\documentclass{aa}  
\usepackage{graphicx}
\usepackage{rotating}
\usepackage{lscape}
\usepackage{longtable}
\usepackage{pdflscape}
\usepackage{subfigure}
\usepackage{hyperref}
\usepackage{txfonts}
\usepackage{color}
\usepackage[dvipsnames]{xcolor}
\usepackage{natbib}
\usepackage{enumerate}
\bibpunct{(}{)}{;}{a}{}{,} 
\usepackage{bibentry} 
\nobibliography*

\begin{document} 

\title{The Gaia-ESO Survey: No sign of multiple stellar populations in open clusters from their sodium and oxygen abundances\thanks{Based on observations collected at the ESO telescopes under programme 188.B3002, 193.B-0936, and 197.B-1074, the {\it Gaia}-ESO Public Spectroscopic Survey.}}


   \author{A. Bragaglia\inst{\ref{INAF-OAS}}
          \and
          V. D'Orazi\inst{\ref{INAF-OAPd},\ref{TorVergata}}
          \and
          L. Magrini\inst{\ref{INAF-OAA}}
          \and
          M. Baratella\inst{\ref{ESOChile}}
          \and
          T. Bensby\inst{\ref{lund}}
          \and
          S.L. Martell\inst{\ref{NSW1},\ref{NSW2}}
          \and
          S. Randich\inst{\ref{INAF-OAA}}
          \and
          G. Tautvai{\v s}ien{\. e}\inst{\ref{Vilnius}}
          \and
          E.~J. Alfaro\inst{\ref{andalusia}}
          \and
          L. Morbidelli\inst{\ref{INAF-OAA}}
          \and
          R. Smiljanic\inst{\ref{rodolfo}}
          \and
          S. Zaggia\inst{\ref{INAF-OAPd}}
          }

   \institute{INAF-Osservatorio di Astrofisica e Scienza dello Spazio di Bologna, via P. Gobetti 93/3, 40129, Bologna, Italy \label{INAF-OAS}\\
              \email{angela.bragaglia@inaf.it}
               \and
              INAF-Osservatorio Astronomico di Padova, vicolo dell'Osservatorio 5, 35122, Padova, Italy \label{INAF-OAPd}
              \and
               Dipartimento di Fisica,  Universit\`a degli Studi di Roma Tor Vergata, via della Ricerca Scientifica 1, 00133, Roma, Italy \label{TorVergata}
             \and
              INAF-Osservatorio Astrofisico di Arcetri,
              largo E. Fermi 5, 50125, Firenze, Italy \label{INAF-OAA}
              \and
              European Southern Observatory, 
              Alonso de C\'ordova 3107,
              Vitacura, Santiago, Chile
              \label{ESOChile}
              \and
	      Lund Observatory, Division of Astrophysics, Department of Physics,
              Lund University, Box 118, SE-22100 Lund, Sweden \label{lund}
              \and
              School of Physics, University of New South Wales, Sydney, NSW 2052, Australia \label{NSW1}
              \and
              Centre of Excellence for Astrophysics in Three Dimensions (ASTRO-3D), Australia \label{NSW2}
             \and
              Institute of Theoretical Physics and Astronomy, Vilnius University, Saul\.{e}tekio av. 3, LT-10257 Vilnius, Lithuania \label{Vilnius}
              \and
              Instituto de Astrof\'{i}sica de Andaluc\'{i}a-CSIC, Apdo. 3004, 18080, Granada, Spain \label{andalusia}
              \and
              Nicolaus Copernicus Astronomical Center, Polish Academy of Sciences, ul. Bartycka 18, 00-716, Warsaw, Poland \label{rodolfo}
              }

 
  \abstract
  {The light element (anti-)correlations shown by globular clusters (GCs) are the
main spectroscopic signature of multiple stellar populations. These internal
abundance variations provide us with fundamental constraints on the formation
mechanism of stellar clusters.}
  {Using $Gaia$-ESO, the largest and most homogeneous survey of open clusters
(OCs), we intend to check whether these stellar aggregates display the same
patterns. Based on previous studies of many GCs, several young and massive
clusters in the Magellanic Clouds, as well as a few OCs, we do not expect to find any
anti-correlation, given the low mass of Milky Way OCs.  }
  {We used the results based on UVES spectra of stars in $Gaia$-ESO to derive the
distribution of Na and O abundances and see whether they show an unexplained
dispersion or whether they are anti-correlated. By selecting only high-probability members with
high-precision stellar parameters, we ended up with more than 700 stars in 74 OCs.
We examined the O-Na distribution in 28 OCs with at least 4 stars available as well as
the Na distribution in 24 OCs, with at least 10 stars available.}
  {We find that the distribution of Na abundances is compatible with a
  single-value population, within the errors. The few apparent exceptions can be
  explained by differences in the evolutionary phase (main sequence and giant post
  first dredge-up episode) or by difficulties in analysing low gravity giants. We
  did not find any indication of an Na-O anti-correlation in any of the clusters for
  which O has been derived.}
 {Based on the very small spread we find, OCs maintain the status of single
stellar populations. However, a definitive answer requires studying more elements
and larger samples covering different evolutionary phases. This
 will be possible with the next generation of large surveys.}

\authorrunning{Bragaglia et al.}
\titlerunning{Na-O in {\it Gaia}-ESO OCs}
   \maketitle
%

\section{Introduction \label{intro}}

Stellar clusters could serve as the best exemplification of a simple stellar
population, since they are composed of stars of different masses that were born
together and characterised by the same age and initial chemical composition.
Thus, they offer an optimal way to study stellar and galactic evolution,
representing ideal benchmarks. However, this simple view \citep[see][to cite
only one historical review]{renzini_fusipecci} had to be abandoned in the case of
globular clusters (GCs). Starting from low-resolution spectroscopy indicating
anti-correlated star-to-star variation in CN and CH \citep[e.g.][]{smith93} and
from the high-resolution works by the Lick-Texas group \citep[e.g.][]{kraft93}, 
evidence has mounted to suggest that stars in GCs display light element
abundances (in particular, O, Na, Mg, and Al), which are vastly different from
those of field stars of similar metallicity \citep[see e.g.][for low-metallicity
field stars and GCs]{gratton00,gratton01}. The new paradigm for GCs is that they
are composed by multiple populations, as described, for instance, in recent
reviews by \citet{gratton12,gratton19,bastian_lardo}. This is  important when
seeking to understand how GCs form. The question of whether 'multiple'  also
points to different ages (as suggested by most models of early chemical
enrichment) is still a matter of debate.  What is clear is that stars of very
different chemical compositions (with respect to  proton-capture elements and
helium) coexist in the same GC. This is amply seen in photometry as well,
especially when combinations of filters sensitive to light element variations
are used \citep[as in the works using Johnson, Str\"omgren, or Hubble Space
Telescope UV filters, see e.g.][]{monelli13,carretta11,milone17}. Initially,
only GCs in the Milky Way (MW, all very old, see e.g. \citealt{kruijssen19}) had
been studied in the past, followed by massive clusters in nearby galaxies, such
as the Magellanic Clouds later on \citep[e.g.][]{mucciarelli09}. Light-element
variations were searched for via spectroscopic or photometric means and were
indeed found in (massive) clusters down to an age of 2 Gyr
\citep[e.g.][]{martocchia19,shen_oh23}.  

Interestingly, light-element (anti-)correlations have been found for all MW GCs
examined \citep{bastian_lardo,gratton19}, possibly with the exception of
Ruprecht 106 \citep{rup106}; however, this is not the case for open clusters
(OCs). These younger, less massive, and more metal-rich stellar clusters have
not been extensively studied in this context. As far as we know, only two old
and massive OCs (i.e. Berkeley~39 and NGC~6791) have been observed on purpose
with high-resolution spectroscopy to see whether they host multiple populations.
\citet{bragaglia12,bragaglia14} and \citet{villanova18} examined some tens of
giant stars in these two OCs, looking for variations is light elements, but did
not find any evidence of them. Literature studies on OCs do not usually have
large samples and are not focussed on this subject, but generally no evidence
has been found either, as seen in the discussion in \citet{maclean15}. There is
one exception, however: \citealt{pancino18} collected literature data on four
nearby intermediate-age OCs and found indication of variations in
Na, O, and Mg among main sequence stars that are similar to those seen for GCs.
None of these four OCs has a mass comparable to that of GCs, indicating
that we cannot safely exclude OCs only because of their low mass.  The features
are enhanced in fast-rotating massive stars ($v \sin~i>50$ km~s$^{-1}$).  These
authors noted that these anomalous abundances would not survive the first
dredge-up, thus disappearing in giants. 

Indeed, the first dredge-up plays a role in Na abundances, which may be
increased after it in young clusters.  Stellar evolutionary models, for instance
by \citet{lagarde12}, indicate an enhancement in stars with mass larger than
about 2 M$_\odot$ at the main sequence turn-off. This has been confirmed by
observations of open clusters and Cepheids, both within GES and in additional
samples, see \cite{smiljanic16,smiljanic18} for more details. This point will be
discussed later in our paper for the few OCs with both dwarf and giant stars
observed.

All this calls for further analysis, involving both dwarf and giant stars.
Luckily,  the paucity of studies on multiple populations in OCs may now be
amended using the large data set produced by {\it Gaia}-ESO (GES hereinafter).
{\it Gaia}-ESO is a public spectroscopic survey conducted with FLAMES at the ESO
VLT\footnote{ESO stands for European Southern Observatory, VLT for Very Large
Telescope, FLAMES for Fibre Large Array Multi Element Spectrograph (described in
\citealt{pasquini02}). FLAMES has two spectrographs, UVES (Ultraviolet and
Visual Echelle Spectrograph, \citealt{dekker00}) at resolution $R\sim45000$ and
GIRAFFE, at intermediate resolution ($R\sim20000$, depending on the setup).} 
over 340 nights from December 2011 to January 2018. A full description of the
survey goals and strategy can be found in \cite{gilmore22,randich22}.
Briefly, GES observed about 115000 stars in the thin and thick disc, bulge, and
halo of the Milky Way. It dedicated a large fraction \citep[about 37\%,
see][]{randich22} of the time to OCs and star-forming regions. For a description
of goals, main results, and selection process, we refer to
\citet{randich22,bragaglia22}. In total, 62 stellar clusters were observed by
GES for science and a handful for calibration \citep{pancino17}. Then, data for
18 further OCs were retrieved from the ESO archive (mostly UVES spectra). A
complete list with main properties can be found in \citet{randich22} and in
Table~\ref{infoA1}. In each cluster, GES obtained spectra of stars in all
evolutionary phases (from about 100 to more than 1000 stars).
\citet{bragaglia22} describes the cluster types, the observational strategy, and
the kind of targets observed.

All spectra are available from the ESO archive (both raw and science-ready
spectra) online \footnote{ 
\hyperlink{http://archive.eso.org/cms.html}{http://archive.eso.org/cms.html}}. A
catalogue containing radial and projected rotational velocities, stellar
parameters (effective temperature, surface gravity and metallicity), abundances
of many elements, parameters for tracing accretion and activity in young stars,
etc. can be found at the ESO Catalogue website.\footnote{
(\hyperlink{https://www.eso.org/qi/catalogQuery/index/393}{https://www.eso.org/qi/catalogQuery/index/393})}

In the present paper, we describe the data used in Sect.~\ref{data}, along with
the the Na and O distributions in Sect.~\ref{nao}, where we also discuss some
clusters. A summary and conclusion are provided in Sect. \ref{summary},  while
additional information on the clusters and individual targets is given in
Appendix.

\section{The data \label{data}}

We used the GES final, public catalogue and selected only the observations in
the OC fields, using the field {\tt GES\_FLD,} which contains the name of the
clusters. We then cross-matched the selected stars with {\it Gaia} EDR3
\citep{gaiaEDR3} using the TOPCAT table access protocol and a search radius of 2
arcsec. After the $Gaia$ DR3 release \citep{gaiaDR3}, we added also the radial
velocity (RV) information. The match resulted in 42776 stars, but only part of
them actually cluster members, as reported, for instance, in \citet{bragaglia22}
and \citet{jackson22} for some statistics on membership. To select only high
probability members we used \citet[][based on {\it Gaia} DR2
astrometry]{tristan20} and \citet[][which also includes information from GES,
such as RV, temperature, and so on in addition to {\it Gaia} EDR3
astrometry]{jackson22}. Keeping only stars with a probability larger than 0.7 in
at least one of the two studies, we obtained about 14000 stars.

As we were interested in the abundances of Na and O,  we further selected only
stars observed with UVES (at a resolution of 47000). In particular, we kept only
stars observed with the U580 setup, covering the 4800-6000 \AA \, region. The
main GIRAFFE setups used for the OCs  \citep[HR15n and HR09b, see
][]{bragaglia22,randich22}  do not contain lines of either element\footnote{We
did not consider Mg and Al, also available  from the GIRAFFE spectra, because
they are not known to vary as much as O, Na in metal-rich GCs. Indeed, the
production of Al is larger in the metal-poorer and more massive GCs, as first
noted by \citet{carretta09b} and later confirmed on larger samples, a
metallicity that is well below and a mass well above that of OCs.}. This
selection drastically reduced our sample to 970 stars.  Finally, we applied
quality cuts similar to what was done in other GES papers and kept only stars
for which errors on temperature, gravity, and metallicity were less than 100~K,
0.3 dex, and 0.1 dex, respectively (all three conditions met). 

The only available O lines were the [O {\sc i}] transitions at 6300 and 6363
\AA, more easily measured in giants than in dwarfs. Correction for telluric
lines was not part of the standard GES processing \citep[see][for the UVES
spectra]{sacco14}. This means that oxygen abundance was measured from the [O
{\sc i}] 6300 \AA \, line only in part of the clusters when the combination of
intrinsic line of sight (LoS) velocity  and barycentric motion kept the line
free from contamination. We indicate in Table~\ref{infoA1} the cases for which O
was measured and for which we could hope to explore the existence of a Na-O
anti-correlation (however, see the limitations imposed on precision and the
minimum number of stars; for instance, in IC~2602 we have 18 stars, but O was
measured only in one).

A detailed description of how UVES spectra are reduced and stellar parameters
and abundances are computed in GES can be found in
\citet{sacco14,smiljanic14,randich22,gilmore22}. Briefly, after data reduction,
the spectra are analysed by multiple pipelines and astrophysical parameters are
internally homogenised within each working group (WG; here, WG~11, dealing with
UVES spectra). A second homogenisation process, involving results of all WGs and
based on a set of calibrators \citep[benchmark stars, OCs, and GCs;
see][]{pancinocalib} produces the set of recommended parameters published in the
released catalogue. Abundances from UVES spectra are then computed by each
analysis node with the homogenised stellar parameters. The abundances from the
different nodes are combined line by line with a Bayesian approach by WG~11 (see \citealt{worley24}
for the updated description of the WG~11
approach). The abundances of O, C, and N (the last two from molecular bands) are
an exception since they are measured only by the Vilnius analysis node. 
Finally, the final abundances are released by WG~15, dedicated to quality
control, homogenisation, and preparation of the final catalogue. 

Abundances in GES were calculated assuming local thermodynamic equilibrium (LTE)
approximation. Ideally, a correction should be applied to non-LTE (NLTE), which
offers a better description. Indeed, NLTE corrections depend on many factors (in
particular on temperature and gravity) and avoiding them may skew the
distribution of abundances. However, given the way the GES abundances are
calculated (see \citealt{smiljanic14,gilmore22,randich22}), this is complicated
and we keep to the LTE values. This is irrelevant for O, measured essentially
only in giants (most are red clump, RC, stars) from a forbidden line, that is,
not subject to NLTE effects \citep{asplund04}. For Na, we minimise the effects
by separating dwarfs and giants, thus considering samples with less disparate
parameters.

The final sample contains 735 stars in 84 clusters, 74 of which have at least
three stars (the number of stars surviving all cuts varies from one to 68).  We
give information on stellar parameters, RV, and selected abundances for the
total selected sample in Table~\ref{infoA2}. We have both O and Na abundances
for 270 stars; 240 of them are in 28 OCs with at least 4 stars available, while
22 OCs have only one to three stars. Sodium is measured essentially in all stars
and there are 24 OCs with at least ten stars with Na abundance.

\begin{figure*}
    \centering
    \includegraphics[scale=0.9]{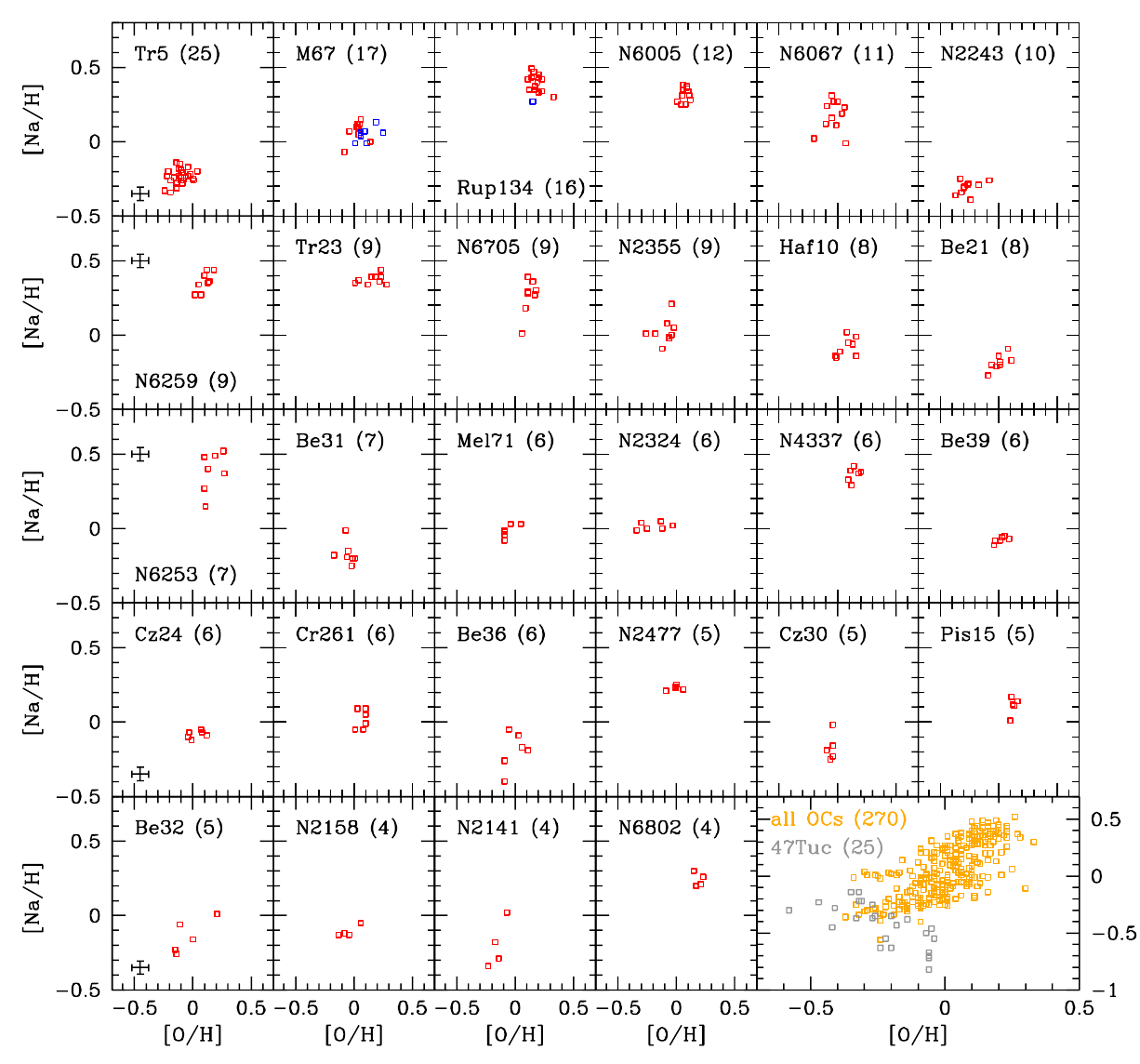}
    \caption{Distribution of [Na/H] versus [O/H] for all OCs with at least four
stars with valid values. The blue points for two clusters indicate dwarf stars,
while all red points are for giants. The bottom right panel displays the available
values in all OCs (in orange) compared to the GC 47~Tuc (grey points). Average
error bars are displayed only in the first column of the plots.}
    \label{fig:nao}
\end{figure*}

\section{The Na and O distribution in GES OCs \label{nao}}

\begin{table*}
\centering
\caption{Information on mean [Na/H], dispersion and IQR([Na/H]) for GES OCs and 47 Tuc. }
\begin{tabular}{lllcrrrrc} 
\hline\hline
Ord &Cluster& Age  &Num &  mean & sigma  &median &median &IQR        \\
    &       &(Gyr) &    &[Na/H] &        &[Na/H] &[Fe/H] &[Na/H]          \\
\hline
1 & NGC 6253     & 3.24 & 68 & 0.478 & 0.099 &    0.38 &    0.36 & 0.13 \\ 
2 & M 67         & 4.27 & 38 & 0.040 & 0.054 &    0.05 & $-$0.02 & 0.10 \\
3 & NGC 3532     & 0.40 & 36 & 0.000 & 0.078 & $-$0.02 & $-$0.01 & 0.12 \\
4 & Trumpler 5   & 4.27 & 26 &-0.245 & 0.038 & $-$0.23 & $-$0.35 & 0.06 \\
5 & NGC 2141     & 1.86 & 23 & 0.004 & 0.064 &    0.01 & $-$0.05 & 0.09 \\
6 & Ruprecht 134 & 1.66 & 20 & 0.399 & 0.045 &    0.41 &    0.27 & 0.09 \\
7 & Trumpler 20  & 1.86 & 26 & 0.184 & 0.041 &    0.17 &    0.13 & 0.08 \\
8 & NGC 2243     & 4.37 & 18 &-0.279 & 0.041 & $-$0.28 & $-$0.47 & 0.06 \\
9 & Berkeley 32  & 4.90 & 17 &-0.141 & 0.085 & $-$0.16 & $-$0.29 & 0.12 \\
10 & NGC 2516    & 0.24 & 15 &-0.039 & 0.051 & $-$0.05 & $-$0.05 & 0.08 \\
11 & NGC 2158    & 1.55 & 14 &-0.076 & 0.038 & $-$0.05 & $-$0.16 & 0.09 \\
12 & NGC 2425    & 2.40 & 14 &-0.059 & 0.066 & $-$0.07 & $-$0.14 & 0.12 \\       
13 & NGC 6067    & 0.13 & 13 & 0.216 & 0.141 &    0.23 & $-$0.03 & 0.15 \\   
14 & Berkeley 21 & 2.14 & 13 &-0.204 & 0.068 & $-$0.20 & $-$0.21 & 0.06 \\
15 & NGC 6005    & 1.26 & 13 & 0.343 & 0.018 &    0.31 &    0.22 & 0.08 \\
16 & Berkeley 81 & 1.15 & 13 & 0.348 & 0.057 &    0.36 &    0.25 & 0.14 \\
17 & NGC 6259    & 0.34 & 12 & 0.379 & 0.045 &    0.38 &    0.17 & 0.10 \\
18 & Haffner 10  & 3.80 & 12 &-0.079 & 0.047 & $-$0.07 & $-$0.12 & 0.10 \\
19 & NGC 2477    & 1.12 & 11 & 0.156 & 0.072 &    0.21 &    0.13 & 0.15 \\
20 & Trumpler 23 & 0.71 & 11 & 0.346 & 0.079 &    0.37 &    0.21 & 0.04 \\
21 & Blanco 1    & 0.10 & 11 &-0.083 & 0.065 & $-$0.07 & $-$0.05 & 0.09 \\
22 & NGC 2420    & 1.74 & 37 &-0.098 & 0.088 & $-$0.08 & $-$0.16 & 0.11 \\
23 & NGC 6705    & 0.31 & 10 & 0.282 & 0.023 &    0.28 &    0.06 & 0.04 \\
24 & NGC 6802    & 0.66 & 10 & 0.269 & 0.029 &    0.27 &    0.15 & 0.06 \\

25 & 47 Tuc (GC) &      & 66 &$-$0.44 & 0.20 & $-$0.43 & $-$0.75 & 0.30 \\
\hline
\end{tabular}
\tablefoot{'Ord' is the index used in Fig.~\ref{fig:iqr} on the x-axis to indicate
the cluster; 'Num' indicates how many stars were used to compute the mean,
dispersion, and median [Na/H]; the last column shows IQR([Na/H]), that is the
difference between the 3rd and 1st quartiles of the [Na/H] distribution.}
\label{t:iqrna}
\end{table*}

\subsection{Considering O and Na}

In Fig.\ref{fig:nao}, we show the distribution of O and Na abundances in the 28
OC, individually and all together. We use [O/H] and [Na/H] to avoid introducing
the effect of [Fe/H], which shows a bland dependence on temperature and gravity
(a discussion, employing OC data, is presented in \citealt{magrini23}). The effect could mask or introduce spreads when stars of very
different parameters are considered in the same cluster (a few cases are
discussed in Sect.~\ref{odd}). However, we show in Appendix
(Sect.~\ref{nafe-ofe}) that using [O/Fe] and [Na/Fe] does not change our
conclusions.

We also compared the distribution(s) to that of the GC 47~Tuc, analysed by GES
in a homogeneous way. This cluster was chosen because it is a metal-rich GC
showing a ''short" anti-correlation (see \citealt{carretta09a,carretta09b} for a
comparison to other globulars). As we see from the right-most lower panel of
Fig.\ref{fig:nao}, there is a different (minimum) value for Na and (maximum)
value for O in the OCs, which are a thin-disk, relatively young population, and
in the GC. This is evident also in Fig.~\ref{app-nao}, where we see that 47~Tuc
has the typical high [$\alpha$/Fe] abundance ratio of the old stellar
populations. In fact the [O/Fe] values of the 47~Tuc stars of primordial
composition (high O and low Na) are larger than for OC stars.

Crucially, we do not see an indication of anti-correlation between O and Na in
any of the OCs  in Fig. \ref{fig:nao}. Either the distributions are compatible
with a dispersion due to errors (in NGC~6005, Tr~23, Be~39, etc.), or we see a
correlation (in NGC~2243, NGC~6259, NGC~6705, etc.), which is however not
significant. Based on  these data, OCs should be considered basically different
from GCs.

However, the evidence against OCs hosting multiple populations may still be
considered not conclusive. First, we can address the issue in about one-third of
the whole GES sample because O was measured only in a fraction of the clusters.
Second, we have only six clusters with at least 10 stars with both abundances,
at variance with the dedicated studies on NGC~6791 and Be~39 mentioned in the
introduction (35 stars in \citealt{bragaglia14} and 17 in \citealt{villanova18}
for the first, and 29 in \citealt{bragaglia12} in the latter). In addition, the
extension of the O-Na anti-correlation shows a dependence on cluster mass in
GCs, see the measure of the inter-quartile ratio (IQR) of [O/Na] (see
\citealt{carretta10, gratton19} for examples). Open clusters have generally
(much) lower masses than GCs (see e.g. Fig.~7 in \citealt{gratton19}); a simple
linear extrapolation of the relation between mass and IQR[O/Na] would lead to
very small (even negative) values. For a direct comparison of total masses, see
for instance \citet{baumgardt} for GCs (updated online\footnote{
\hyperlink{https://people.smp.uq.edu.au/HolgerBaumgardt/globular/}{https://people.smp.uq.edu.au/HolgerBaumgardt/globular/}}),
with values about 10$^3- 10^6$ M$_\odot$ and \citet{ebrahimi22,almeida2023} for
15 OCs and more than 700 OCs, respectively, with a few hundreds to a few
thousands M$_\odot$. In Table~\ref{infoA1} we indicate the masses of the 23 GES
OCs which are present in these two sources.

\begin{figure*} \centering \includegraphics[scale=0.95]{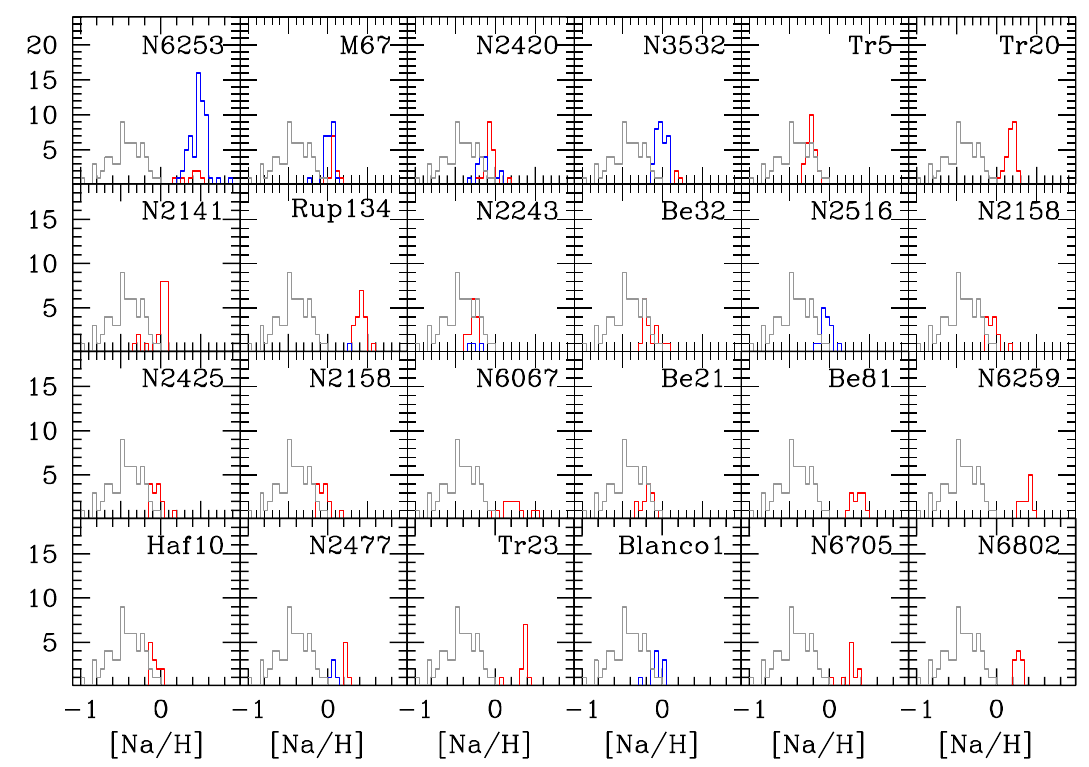}
\caption{Histogram of [Na/H] for all OCs with at least ten stars with valid
values. The blue histograms are drawn for dwarfs and the red histograms for
giants in each cluster, respectively. The histogram for 47~Tuc (grey histogram)
is also shown in each panel for comparison.} \label{fig:histona} \end{figure*}

\begin{figure} \centering
\includegraphics[scale=0.47]{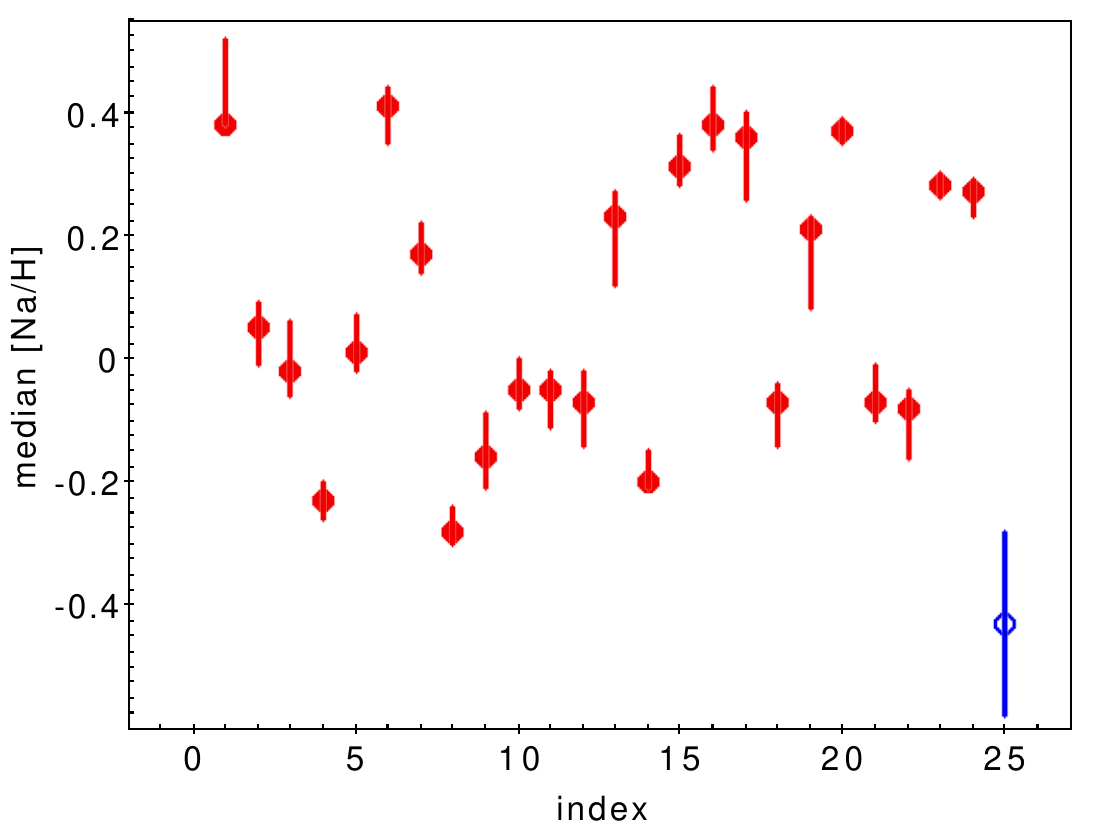}     
\caption{Median value of [Na/H] for  OCs with at least ten valid Na values; the
vertical lines indicate the first and third quartile of the distribution in each
cluster, that is, the IQR for [Na/H]. The rightmost point (blue open circle) is
for 47~Tuc. The value of index identifying each OC can be found in
Table.~\ref{t:iqrna}}
\label{fig:iqr} 
\end{figure}

\begin{figure} \centering \includegraphics[scale=0.45]{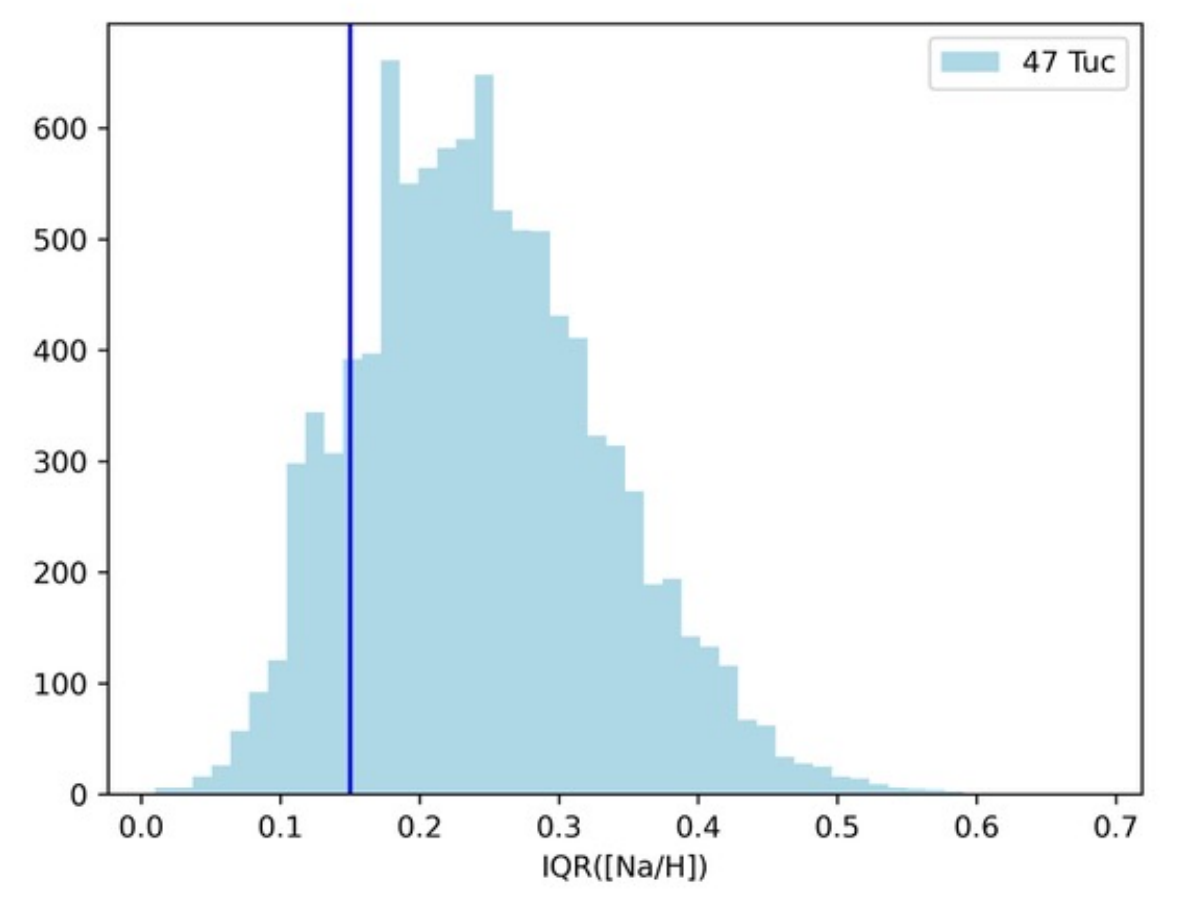}    
\caption{Histogram of the IQR([Na/H]) values for 47~Tuc if only 10 stars are
randomly selected among those included in GES. We did 10000 realisations of the
IQR and we obtain a value larger than the largest for OCs (0.15, indicated by
the vertical line) in 85 \% of the cases.} 
\label{fig:iqr47Tuc} 
\end{figure}

\subsection{Considering only Na \label{onlyNa}}

We can also look at Na alone to check whether a significant star-to-star
variation exists. This permits us to include a few clusters for which no, or
very few, O measures are available (NGC~2420, NGC~3532, NGC~2514, NGC~2425,
Tr~20, Be~81, and Blanco~1 are in the first category, NGC~2141, NGC~2158,
NGC~2477, NGC~6802, and Be~32 in the second). This is apparent from the
comparison of Figs.~\ref{fig:nao} and~\ref{fig:histona}. In the latter, we only
plot the [Na/H] histograms of OCs with at least ten valid measurements. The
analogous plot involving [Na/Fe] is shown in Fig.~\ref{app-histona}. In
Fig.~\ref{fig:histona} we also plot the histogram of [Na/H] values in 47~Tuc in
each panel for immediate comparison: apart from a different zero point (see
Fig.~\ref{app-histona}, where the metallicity is taken into account), the
distribution of all OCs is always much narrower (about two to six times
narrower, judging from Table~\ref{t:iqrna}). For the OCs, we separated dwarfs on
the main sequence (MS) and giants by selecting stars with $\log~g$ larger or
smaller than 3.5, respectively. This allowed us to avoid seeing the effect of
evolution on Na abundance (see Sect.~\ref{odd}) and also to minimise possible
differences due to LTE assumptions (see Sect.~\ref{data}).

Table~\ref{t:iqrna} lists the 24 resulting OCs and gives information on the
number of stars, the mean [Na/H] values together with standard deviations, 
median and the IQR([Na/H]) values.  The mean [Na/H] and its intrinsic dispersion
(cols. 5-6) have been computed using a maximum likelihood algorithm (kindly made
available by the authors), which takes into account also the errors, better
computing the intrinsic dispersion (for details, see \citealt{mucciarelli12}). 
For comparison, we list the same values for 47~Tuc. In Fig.~\ref{fig:iqr},  we
plot the median values for the 24 OCs, together with the extension of the
variation (the lines show the first and third quartile of the [Na/Fe]
distributions). The extensions are always small, see the comparison with 47~Tuc,
also shown in the figure.  As the number of star with Na abundance is
larger than in the OCs, with the exception of NGC~6253, we tested the effect of
having only a small number of stars on the IQR. Figure~\ref{fig:iqr47Tuc} shows
the distribution of IQR[Na/H], when only 10 stars are randomly extracted from
the 47~Tuc sample (we selected 10 as this is the smallest number of stars for
which we measured IQR in GES OCs). The random extraction was repeated 10000
times and we measured an IQR[Na/H] larger than 0.15, that is, the highest value
found for OCs, in 85\% of the times.

Also, when using only the Na abundances, we did not see any indication of large
star-to-star variation. In particular, we do not see any unexpected Na
enhancement (see also Sect.~\ref{odd}) similar to that seen in GCs. No clear
evidence of multiple populations is found, although given the differences in age
and mass, the results are inconclusive as to whether multiple populations are
actually present.

As a sanity check, we considered two non-GES sets of data of comparable
size. First, we took the eleven giant stars in NGC~6705 analysed in
\cite{n6705_apo}; these authors used the  BACCHUS code on APOGEE spectra and
derived abundances of many elements (including Na). The mean and intrinsic
dispersion of [Na/H], computed as for the GES data, are +0.46 (assuming 6.18 as
solar value) and 0.039. While the Na abundance is clearly different from the GES
value, the dispersion is comparable and compatible with their errors. Second, we
took Stock~2, a cluster where both dwarfs and giants were observed by the SPA
large programme at the TNG using the very-high-resolution spectrograph HARPS-N
\citep[see][who estimated an age of about 0.5 Gyr]{alonso-santiago2021}. For
Stock~2, there are Na abundances for 10 giants and 12 main sequence stars. If we
consider them together we have a mean [Na/H] of +0.145 with sigma 0.052, similar
to what we get for GES clusters. When we split the sample between dwarfs and
giants we get averages of 0.217 and 0.07, respectively, with an essentially zero
intrinsic dispersion. While the dispersion is always small, we see that it is
safer to separate dwarfs from giants in clusters young enough to display an
enhanced Na abundance after the first dredge-up. That means clusters with mass
at the turn-off larger than 2 M$_\odot$, according to
\citealt{smiljanic16,smiljanic18}. For the GES OCs in Table~1, this may affect
only NGC~3532 and NGC~2477, both discussed in Sect.~\ref{odd}. 

\begin{figure*}
    \centering
    \includegraphics[scale=0.55]{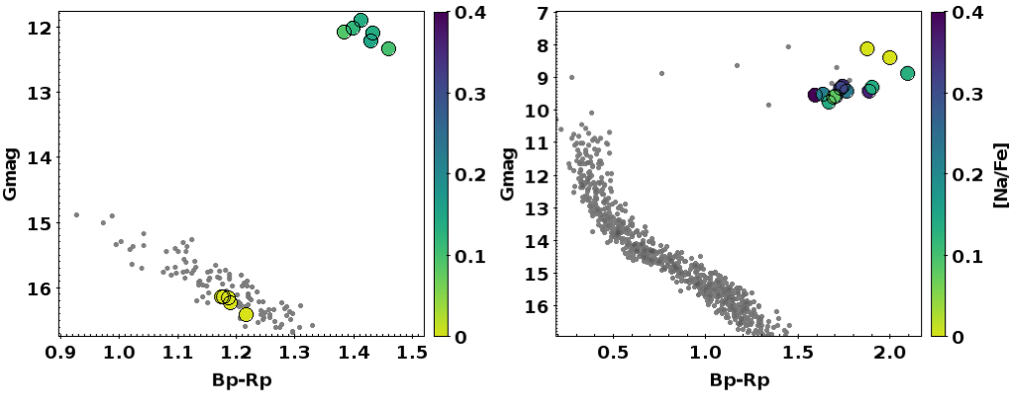}        
    \caption{Colour-magnitude diagrams for NGC~2477 (left) and NGC~6067 (right),
drawn using $Gaia$ photometry of stars observed by GES. Stars with valid Na
abundances are shown as larger symbols, coloured according to their [Na/Fe]
value.}
    \label{fig:cmd_na}
\end{figure*}

\subsection{Some odd cases \label{odd}}

Using the [O/Fe] and [Na/Fe] values, a larger than expected variation in the
O-Na or Na distribution is visible in a few OCs (see Figs.~\ref{app-nao},
\ref{app-histona}, and Table~\ref{infoA2}). In other OCs, only one star deviates
from the bulk distribution in each cluster. \footnote{While checking these
stars, we made use also of the RV in $Gaia$ DR3, finding that many stars are
candidate or confirmed binary systems. We present a comparison of GES and GDR3
velocities, indicating discrepant cases,  in  the appendix
(Sect.\ref{rv-gdr3})}. We discuss  some exemplary cases below.

\paragraph{NGC~2477 -} This is one of the few clusters where both MS and giant
stars were observed. At its age (about 1 Gyr, see Table~\ref{infoA1}), we expect
to see some Na enhancement in giants past the first dredge-up, see
\citet{smiljanic16,smiljanic18}. Also looking at Fig.~\ref{fig:histona}, dwarfs
and giants have separated distribution in [Na/H]. Indeed, as we see from
Fig.~\ref{fig:cmd_na}, left panel, the stars with higher Na content are all
located on the  RC, and those with lower Na are all on the MS (as in the
case of Stock~2, discussed above). Stars within each group have very similar Na
abundance, while there is an offset of about 0.2 dex in [Na/Fe] between MS and
RC stars. For NGC~2477, the larger Na dispersion is then due to an evolutionary
phenomenon. Note that in older clusters, such as M67 and NGC~2243, where we do
not expect Na mixing at the first dredge-up, we do not see a difference in the
Na abundances of MS and giant stars.

\paragraph{NGC~3532 -} The cluster has an age of $\approx400$ Myr and also in
this case both MS and giant stars were observed (34 and three valid [Na/Fe]
values, respectively). Again, the average [Na/Fe] abundances differ by about 0.2
dex, with giants having larger Na abundance. We conclude that also in this case
we are looking at an evolutionary effect (which we should expect whenever the
cluster is younger than about 1 Gyr and has a turn-off mass larger than about 2
M$_\odot$; see Fig.~2 in \citealt{smiljanic18})

\paragraph{NGC~6067 -} In this cluster (age 125 Myr), all stars observed with
UVES are giants (see right panel of Fig.~\ref{fig:cmd_na}). However, given the
young age, all of them have very low gravity ($\log~g$ from 0.7 to 1.8) The
analysis of low-gravity stars is problematic, as discussed at length in the GES
paper dealing with abundance gradients based on OCs (\citealt{magrini23}, to
which we refer to find other OCs and stars affected).  In this case, the larger
dispersion in Na (confirmed also for other elements, such as Al, Mg, Si, Ca, Ti
{\sc i}, at 0.17, 0.10, 0.14, 0.14, and 0.21 dex, respectively) seems to be due
to analysis issues and not to an intrinsic difference in abundance values. 

\paragraph{Individual cases -} There are some stars with high [Na/Fe] compared
to the cluster average. One is in Mel~71  ($Gaia$ DR3 ID 3033958747506151424),
which has [Na/Fe]=+0.44 (compared to an average value of +0.11  of the other
stars, while O is not measured. However, its Na abundance seems in line with the
other elements, the effect is due to its low metallicity ([Fe/H]=-0.39, compared
to -0.12). The star is a binary \citep[it is MMU 29 in][]{mermilliod08}, also
confirmed by the RV difference between GES and GDR3. Another one is in M67
($Gaia$ DR3 604922985178465152). In this case, [Na/Fe]=+0.30 (compared to an
average value of +0.06 without it) and again the star has a Na abundance similar
to the others, no measurement of O, a lower than average metallicity, and is a
binary based on the RV difference between GES and GDR3. In both cases, the stars
are very high probability astrometric members.

There are also stars with [Na/Fe] values much lower than the cluster average.
Two cases are star $Gaia$ DR3 3029944366134389888 in NGC~2425 and star $Gaia$
DR3 2893944295419313280 in NGC~2243 (neither has O measured). The first is a
low-gravity star (see the caveat on the difficulties in their analysis) and is
probably a binary, the second has a discrepant metallicity ([Fe/H]=--0.15 dex,
higher than the cluster average, see Table\ref{t:iqrna}). Finally, the lowest
[Na/Fe] star in NGC~6067 ($Gaia$ DR3 5932570607366776960),  actually has the
highest [Na/H]; however, although the star passed the quality cuts and is a
member both according to its RV and astrometry, its [Fe/H], at +0.5 dex compared
to a median of $-0.03$ dex (Table~\ref{nao}), indicates possible problems in the
analysis. 

In conclusion, these high-probability cluster members seem to be a mixed bag of
discrepant objects. In some cases, we are dealing with binaries and their
metallicities and abundances may be lower due to veiling from a secondary
component. In others, their low gravity may explain the discrepancies. 
While they surely merit being checked, further attention is not required for the
main goal of our paper.

\section{Summary and conclusions \label{summary}}

We used the public survey GES to study the possible presence of star-to-star
variations in the light elements O and Na. These elements are known to vary,
even to large extent, more than 1 dex, in GCs. This is possibly the main
chemical signature of multiple populations and is visible only in massive and
old stellar clusters. The lower limit in age is still debated; using results
mainly based on nitrogen variations, which have an effect detectable also with
photometry, there seems to be a convergence on 2 Gyr \citep[see for instance][]{
martocchia19}. However,  \citealt{cadelano22} found photometric
indications of multiple populations on the MS of a 1.5 Gyr old clusters (but,
interestingly, not among the red giants, possibly due to mixing). This indicates
that the coordinated variations between light elements are not limited to the
early Universe conditions.  Mass and age seem to play an important role in
determining whether multiple populations appear and it is not simple to
disentangle their effects, meaning that it is interesting studying (also)
younger and less massive clusters to constrain models. In summary, light
elements variations are present in all MW GCs studied to date (with the 
possible exception of Rup~106) and in many massive clusters in the Magellanic
Clouds and beyond. On the contrary, they have not been found in OCs
(\citealt{maclean15}, see \citealt{pancino18} for an alternative view) when
studied with large enough samples  
to understand whether an extended star formation process is
required.

Keeping in mind the complications due to the different mass and age regimes
of GCs and OCs, we studied the second in this work. $Gaia$-ESO is presently the
best available source of data since it contains a large number of OCs (more than
80, combining GES proper and archive spectra) and large samples of stars in each
of them. Unfortunately, Na and O are available only for the stars observed with
UVES, since no lines of these elements are present in the GIRAFFE setups
employed. Despite this limitation, GES constitutes the largest database we can
tap for our study.  

We used the public, final data release (see Sect.~\ref{intro} for the location
in the ESO catalogue archive). We selected stars belonging to OCs according to
literature astrometry. After a further selection on errors, we ended up with 735
stars in 84 OCs (only 74 with at least three stars). Only part of the OCs have
both Na and O abundance available, all have at least Na. 

We then checked the Na-O distribution, finding no indication of an
anti-correlation similar to that seen in GCs (we used the GC 47~Tuc as a
comparison, as it has been analysed homogeneously by GES). Either the dispersion
is compatible with being due to errors, or we see a correlation, contrary to
what we see in GCs. We also checked the distribution of Na abundances alone,
which was available for more stars and for different clusters. Also in this
case, no unexplained dispersion was found. The few cases showing larger
differences can be explained by evolutionary effects (i.e. differences between
dwarfs and giants post first dredge-up) or by analysis problems.

Based on these results, OCs are simple stellar populations. Further steps to
fully ascertain this, or finding that also these stellar clusters show
`anomalous' abundances and have populations distinct by their chemistry,
comprise the following. Firstly, collecting larger samples of OCs where all
light elements involved in (anti-)correlations in GCs can be measured, that is
C, N, O, Na, Al, and Mg.  This will be possible with large surveys such as WEAVE
and 4MOST, where OCs are part of the scientific targets 
\citep[see][]{weave,4most}, but also GALAH and APOGEE (and its successor SDSS
V) can contribute, as seen from the example at the end of
Sect.~\ref{onlyNa}. Secondly, collecting a large sample of stars in each
cluster, as GES did. Thirdly, analysing samples in the most precise and
homogeneous way possible, taking into account also evolutionary processes
(diffusion, mixing, etc). The next step is checking the possible influence
of rotation and binarity on the derived abundances; and, finally, taking into
account  departure from LTE for all elements involved, in particular Na.

More data on stellar clusters of all kinds will hopefully come from large
surveys that are due to start soon, namely, WEAVE and 4MOST, to fully understand
the nature of OCs. Both surveys have a low-resolution mode (about 5000) 
essentially covering the optical part of the spectra and a high-resolution mode
(about 20000) covering three broad wavelength ranges. The latter mode, in
particular, contains lines of O  (the forbidden  lines) and Na.

\begin{acknowledgements}
AB thanks Eugenio Carretta for useful comments on the manuscript. We thank the
referee for useful comments which improved the paper presentation. This research
has made use of the services of the ESO Science Archive Facility. This research
has made use of the SIMBAD database \citep{wenger}, operated at CDS, Strasbourg,
France and of the VizieR catalogue access tool, CDS, Strasbourg, France (DOI:
10.26093/cds/vizier). The original description of the VizieR service was published
in \citet{vizierori}. This research has made use of NASA's Astrophysics Data
System. We made extensive use of TOPCAT (http://www.starlink.ac.uk/topcat/,
\citealt{topcat}). This work has made use of data from the European Space Agency
(ESA) mission {\it Gaia} (https://www.cosmos.esa.int/gaia), processed by the {\it
Gaia} Data Processing and Analysis Consortium (DPAC,
\url{https://www.cosmos.esa.int/web/gaia/dpac/consortium}). Funding for the DPAC
has been provided by national institutions, in particular the institutions
participating in the {\it Gaia} Multilateral Agreement. These data products have
been processed by the FLAMES/UVES reduction team at INAF/Osservatorio Astrofisico
di Arcetri.    We acknowledge the support from INAF and Ministero
dell’Istruzione, dell’Universit\`a e della Ricerca (MIUR) in the form of the
grant Premiale VLT 2012 and Premiale 2016 MITiC. TB was funded by grant No.
2018-04857 from The Swedish Research Council 
\end{acknowledgements}

%
%

\bibliographystyle{aa} 
\bibliography{biblio.bib}

\begin{appendix} 
\section{Supplementary information}
We present  some supplementary information on the OCs studied here, on the
individual stars, and on a comparison of RV values with GDR3 results.

\subsection{Literature information on the GES OCs \label{lit-info}}

Table~\ref{infoA1} gives information on the OCs observed by GES, taken from
literature: Nobs is the number of member stars observed with UVES; absorption in
$V$, distance in pc, logAge (A$_V$, dist, logt) are taken from \citet{tristan20}, 
with a few exceptions (NGC~6530, Cha I, $\rho$ Oph, and $\gamma$ Vel) taken from
\citet{randich22}; average cluster RV, standard error, and number of members used
to derive them come from \citet{tarricq21}; Y/N indicate whether oxygen was
measured in the cluster; masses for part of the OCs are taken from
\citet{ebrahimi22,almeida2023}; finally, some alternative names are given in last
column (the names indicated by [KC2019] refer to \citealt{kc19}).

\subsection{Using [O/Fe] and [Na/Fe] \label{nafe-ofe}}

As commented in the main text, using the iron-scaled abundances risks introducing
some spurious effects due to the existence of trends of [Fe/H] with temperature
and $\log~g$ (the topic is discussed in  \citealt{magrini23}). However, even when
we use [O/Fe] and [Na/Fe] (adopting 8.70 and 6.18 as solar abundances for O and
Na, respectively), results are not changed from what we see for [O/H] and [Na/H]. 

\begin{figure*}[h]
 \centering
 \includegraphics[scale=0.9]{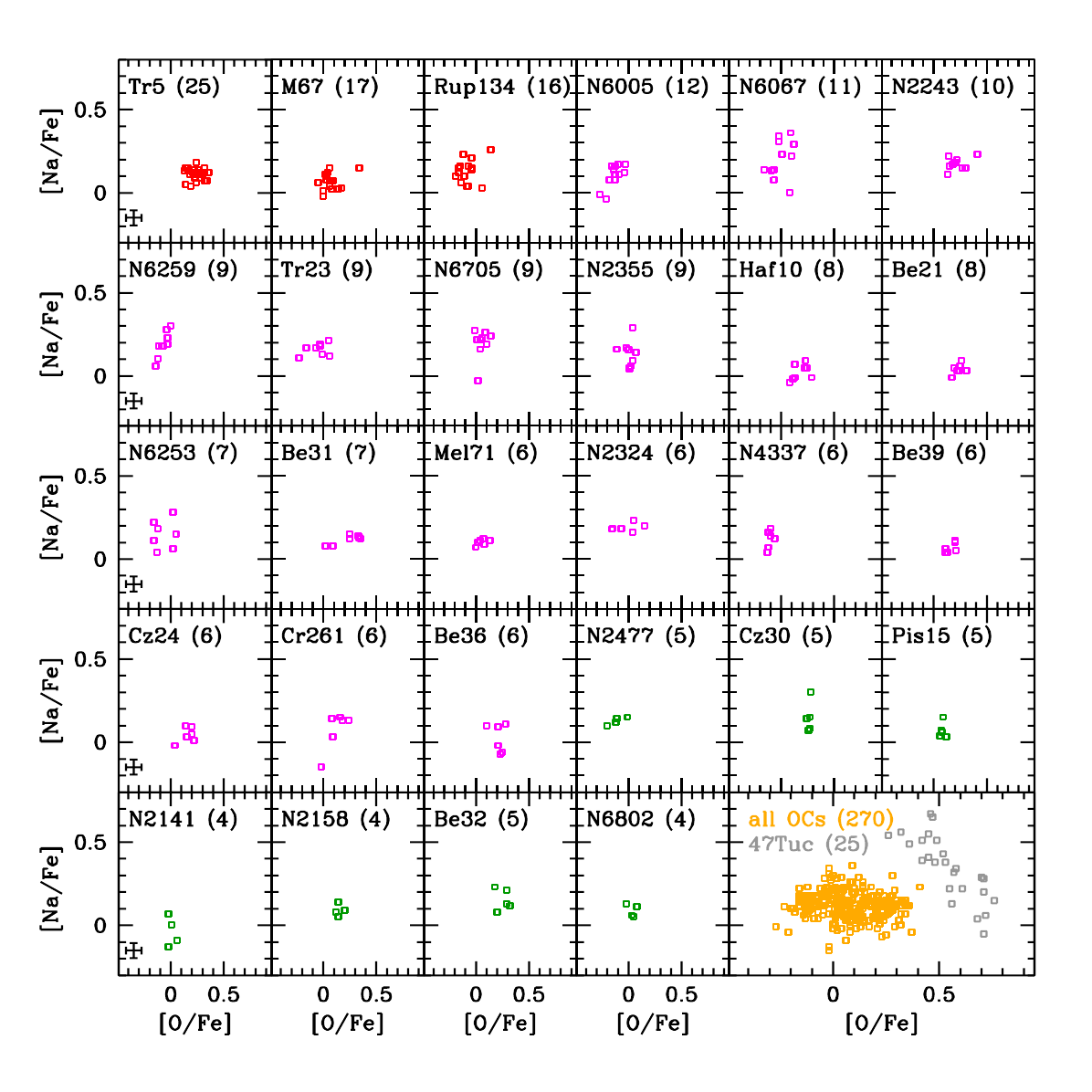}
 \caption{Distribution of [Na/Fe] versus [O/Na] for all OCs with at least four
stars with valid values. The bottom-right panel displays the available values in
all OCs (in orange) compared to the GC NGC~104/47~Tuc (grey points). The colours
indicate clusters with more than 15 stars (red), 5-15 (magenta), and 4 (green).}
 \label{app-nao}
\end{figure*}

Figure~\ref{app-nao} shows the stars in the O-Na plane (analogous to
Fig.~\ref{fig:nao}). In the right-most lower panel we see that essentially all OC
stars have a [Na/Fe] level similar to stars of first generation in 47~Tuc. 

\begin{figure*}[h]
 \centering
 \includegraphics[scale=0.9]{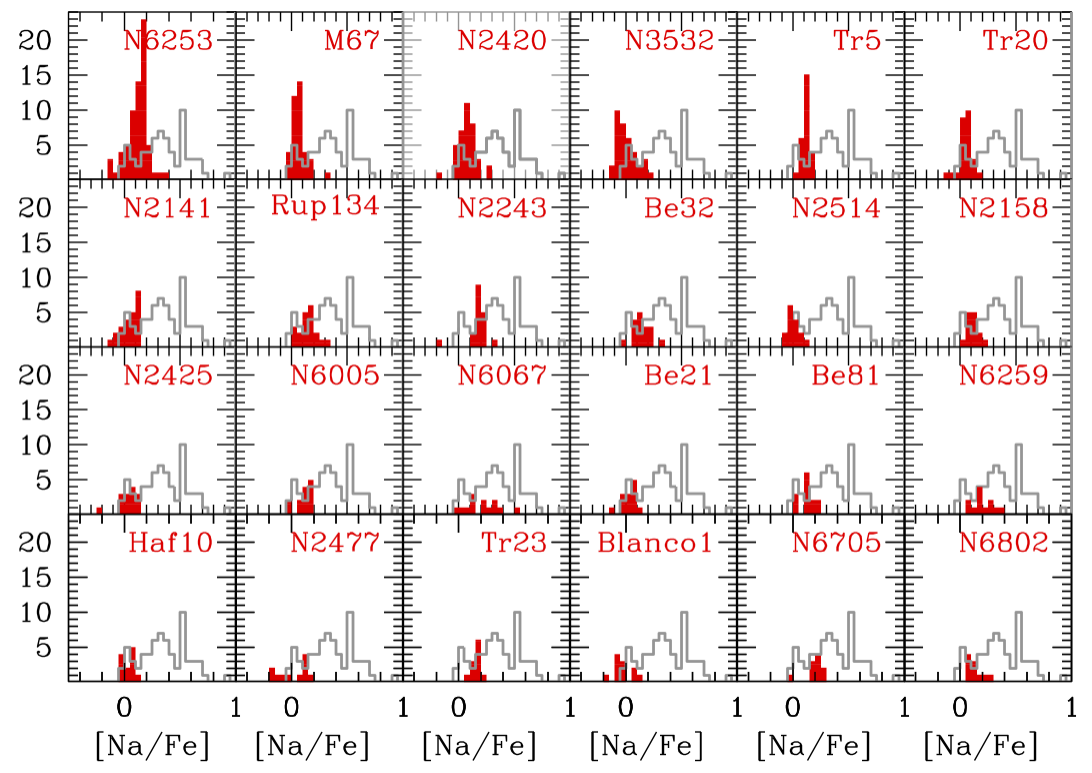}
 \caption{Histogram of [Na/Fe] or all OCs with at least ten stars with valid
values (filled red histograms).The histogram for NGC~104/47~Tuc (grey empty
histogram) is also shown in each panel for comparison.}
 \label{app-histona}
\end{figure*}

Figure~\ref{app-histona} shows the histograms for [Na/Fe] (to be compared to
Fig.~\ref{fig:histona}). Again, we see that OC stars occupy the low-Na part of the
distribution for the GC. 

\begin{figure}[h]
 \centering
 \includegraphics[scale=0.4]{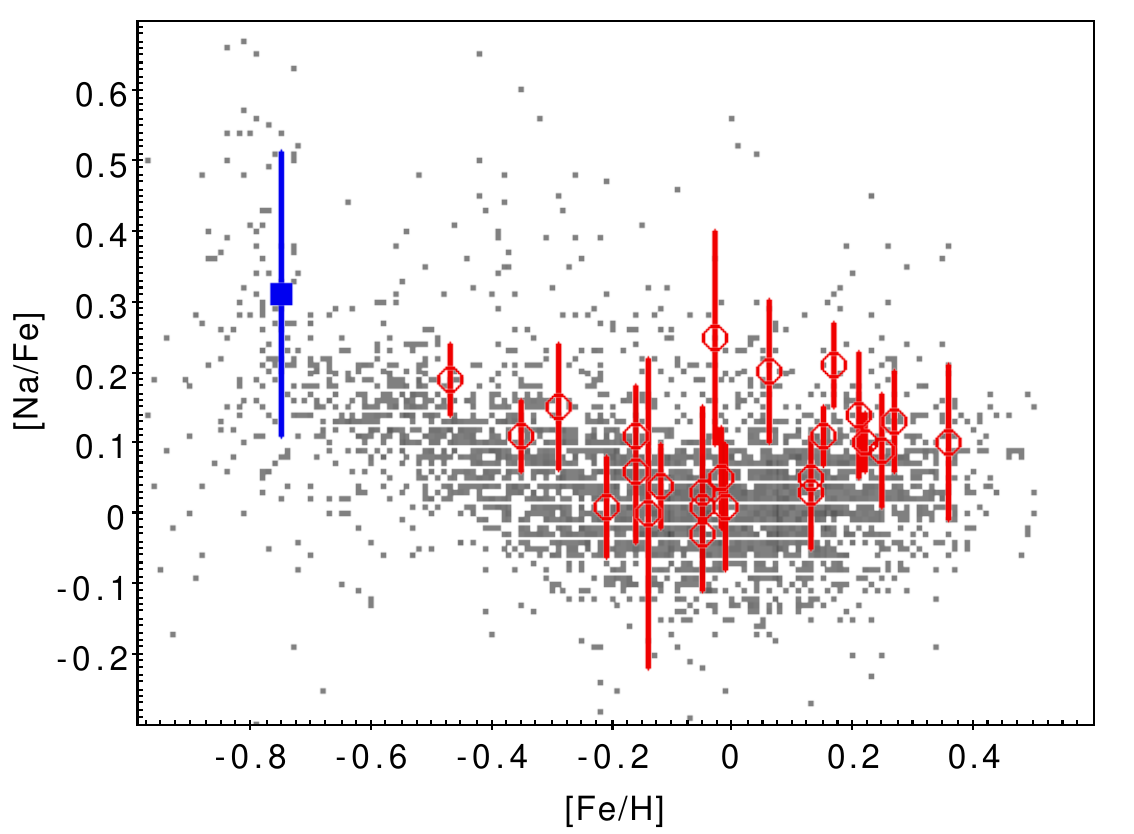}
 \caption{Plot of average [Na/Fe] as function of [Fe/H] for OCs (red circles, with
error bars indicating the standard deviation). The small grey symbols are stars
not in OCs (with the same quality cuts on [Fe/H], T$_{\rm eff}$, and $\log g$
applied to OCs). The GC 47~Tuc is also shown for comparison as a blue filled
square.}
 \label{app-field-oc}
\end{figure}

Finally, Fig.~\ref{app-field-oc} shows the average [Na/Fe] values for the OCs
(with standard deviation indicated by error bars) as a function of metallicity.
This distribution has to be compared to the individual stars in the GES catalogue
and not belonging to OCs, to which we applied the same quality cuts (see
Sect.~\ref{data}). The GC 47~Tuc is shown for comparison, as done throughout the
paper.

\subsection{The individual stars \label{individual}}
Table~\ref{infoA2} gives information on the individual stars, taken from GES and
$Gaia$ DR3. In particular, we list: name of the cluster, $Gaia$ DR3
identification, GES object name, GES RV and error, T$_{\rm eff}$ and error, $\log
g$ and error, metallicity, Na and O abundance with their errors, and $Gaia$ DR3 RV
with error. Only some of the first and last lines are shown here, the complete
table is available at the CDS.

\subsection{Comparison of GES and $Gaia$ RVs \label{rv-gdr3}}
\begin{figure}[h]
    \centering
    \includegraphics[scale=0.45]{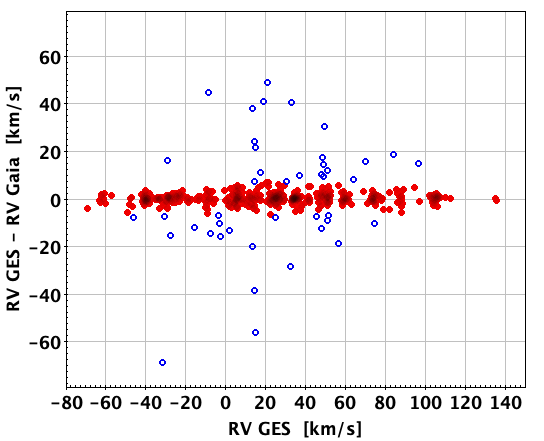}
    \caption{Difference in RV between GES and GDR3 values. The blue open circles
indicate the candidate binary systems.}
    \label{fig:deltaRV}
\end{figure}

We checked the difference between GDR3 and GES velocities. The average difference,
based on 578 stars, is 0.33 ($\sigma$=7.00) km~s$^{-1}$, with values from about
-70 to +50 km~s$^{-1}$. The 535 stars with a $\Delta$RV within 1~$\sigma$ have an
offset of 0.22  ($\sigma$=2.01) km~s$^{-1}$. Figure~\ref{fig:deltaRV} shows the
$\Delta$RV (in the sense GES-GDR3) and the candidate binaries (those with
$\Delta$RV$>1\sigma$) are indicated by  open blue circles.

The 43 candidate binaries are listed in Table~\ref{t:deltaRV}, where (after
the name of the OC), we list the information from GDR3:\ the GDR3 identifier,  RV,
error, and number of transits used for the mean value, Vbroad (a proxy for
rotation velocity) and error when available, and  the renormalised unit
weight error (RUWE); finally, from the GES, RV and error, the membership probability, according to
\citealt{tristan20} and \citealt{jackson22}, and the $\Delta$RV.

We note that besides the significant $\Delta$RV, the large GDR3 errors on Gaia RVS
radial velocity are already a strong indication of binarity. This is especially
true when a large number of transits is available for the stars. For some of the
stars, this is reinforced also by a RUWE larger than 1.4, which is indicative of a
non-single star \cite{gaiaEDR3}; however, the RUWE does not intercept all
binaries.

\onecolumn
\centering
\setlength{\tabcolsep}{1.mm}
\begin{longtable}{lrcrrcrrrcl}
\caption{Literature information on the observed clusters. }\\
\hline
  Cluster &Nobs &A$_V$ &dist &logt  &logMass &RV  &stderr       & N & O?  &Alternative name\\
          &     &(mag) &(pc) &(yr)  &(M$_\odot$) &\multicolumn{2}{c}{(km~s$^{-1}$)} & & & (from Simbad)\\
\hline
\endfirsthead
\caption{continued.}\\
\hline
  Cluster &Nobs &A$_V$ &dist &logt  &logMass &RV  &stderr       & N & O?  &Alternative Name\\
          &     &(mag) &(pc) &(yr)  &(M$_\odot$) &\multicolumn{2}{c}{(km~s$^{-1}$)} & & & (from Simbad)\\
\hline
\endhead
\object{NGC 6253}       & 73  &0.78 & 1653 &9.51 &3.457$^2$           &-24.48 & 0.19 &249 & Y & Mel 156  \\
\object{NGC 3532}       & 43  &0.00 &  498 &8.60 &                    &+5.67  & 0.19 &664 &Y  & [KC2019] Theia 720\\
\object{M67}            & 40  &0.07 &  889 &9.63 &3.266$^2$           &+34.18 & 0.14 &360 & Y & NGC 2682 \\
\object{NGC 2420}       & 37  &0.04 & 2587 &9.24 &                    &+74.78 & 0.09 &326 & N & Cr 154, Mel 69\\
\object{NGC 6705}       & 35  &1.20 & 2203 &8.49 &                    &+34.49 & 0.27 &357 & Y & M11, Cr 391, Mel 213 \\ 
\object{NGC 2547}       & 29  &0.14 &  396 &7.51 &2.559$^2$,2.517$^1$ &+13.05 & 0.16 &134 & N & [KC2019] Theia 74\\
\object{NGC 2516}       & 28  &0.11 &  423 &8.38 &3.296$^1$           &+24.24 & 0.07 &490 & N & [KC2019] Theia 613 \\
\object{NGC 2264}       & 27  &0.79 &  707 &7.44 &2.566$^2$           &+22.54 & 0.26 &141 & Y & [KC2019] Theia 41\\ 
\object{Trumpler 5}     & 26  &1.20 & 3047 &9.63 &                    &-23.10 & 0.19 &  5 & Y & Cr 105\\
\object{Trumpler 20}    & 26  &0.88 & 3392 &9.27 &                    &-39.85 & 0.18 &430 & Y &\\
\object{NGC 2141}       & 23  &0.97 & 5183 &9.27 &                    &+27.11 & 0.23 &429 & Y & Cr 79\\  
\object{Berkeley 39}    & 23  &0.29 & 3968 &9.75 &                    &+59.41 & 0.14 &394 & Y &\\
\object{Ruprecht 134}   & 20  &1.15 & 2252 &9.22 &                    &-39.47 & 0.63 & 72 & Y &\\
\object{NGC 6067}       & 20  &0.97 & 1881 &8.10 &                    &-38.01 & 0.50 &256 & Y & Mel 140\\
\object{NGC 2243}       & 19  &0.02 & 3719 &9.64 &                    &+60.02 & 0.10 &356 & Y & Mel 46\\
\object{IC 2602}        & 18  &0.02 &  149 &7.56 &2.442$^2$,2.236$^1$ &+17.60 & 0.12 & 76 & Y & Cr 229, Mel 102, [KC2019] Theia 92\\
\object{Berkeley 32}    & 17  &0.34 & 3072 &9.69 &                    &+105.68& 0.41 &202 &Y  & Biurakan 8\\
\object{NGC 6530}       & 16  &     & 1325 &6.30 &                    &       &      &    &N  &\\
\object{Blanco 1}       & 16  &0.01 &  240 &8.02 &2.529$^2$,2.594$^1$ &+6.23  & 0.07 &172 & Y \\
\object{NGC 2451A}      &  5  &0.00 &  195 &7.55 &2.534$^2$,2.418$^1$ &+23.42 & 0.13 & 70 & N & [KC2019] Theia 118\\
\object{NGC 2451B}      & 10  &0.18 &  361 &7.61 &2.565$^2$           &+15.23 & 0.09 & 75 & N &\\
\object{IC 2391}        & 14  &0.04 &  148 &7.46 &2.332$^2$,2.207$^1$ &+15.58 & 0.48 & 67 & N & o Vel Cluster, [KC2019] Theia 114\\
\object{NGC 2158}       & 14  &1.44 & 4298 &9.19 &                    &+27.15 & 0.18 &284 & Y & Mel 40\\
\object{NGC 2425}       & 14  &0.89 & 3576 &9.38 &                    &+102.36& 0.60 &132 & Y &\\
\object{NGC 6005}       & 13  &1.42 & 2383 &9.10 &                    &-22.65 & 0.75 & 86 & Y & Mel 138\\
\object{NGC 6259}       & 13  &1.87 & 2314 &8.43 &                    &-32.12 & 1.20 &125 & Y &\\
\object{Berkeley 21}    & 13  &1.96 & 6417 &9.33 &                    &+4.84  & 1.45 & 57 & Y &\\
\object{Berkeley 81}    & 13  &2.75 & 3313 &9.06 &3.295$^2$           &+45.92 & 0.99 & 64 & Y &\\
\object{IC 4665}        & 12  &0.45 &  354 &7.52 &2.489$^2$           &-13.31 & 0.37 & 47 & N & Cr 349, Mel 179, [KC2019] Theia 76\\
\object{Chamaleon I}    & 12  &     &  189 &6.20 &                    &       &      &    & N &\\
\object{Haffner 10}     & 12  &1.35 & 3409 &9.58 &                    &+87.97 & 0.32 & 87 & Y &\\
\object{NGC 2477}       & 11  &0.68 & 1442 &9.05 &                    &+8.46  & 0.10 &197 & Y & Mel 78\\
\object{Trumpler 14}    & 11  &1.00 & 2290 &7.80 &3.300$^2$           &-10.01 & 2.53 & 79 & N &\\
\object{Trumpler 23}    & 11  &2.18 & 2590 &8.85 &                    &-63.75 & 1.57 & 79 & Y &\\
\object{NGC 6802}       & 10  &2.37 & 2753 &8.82 &                    &+10.31 & 1.30 &100 & Y & Cr 400\\
\object{NGC 2355}       &  9  &0.59 & 1941 &9.00 &2.848$^2$           &+36.37 & 0.26 &118 & Y & Cr 133, Mel 63\\
\object{Melotte 71}     &  9  &0.38 & 2139 &8.99 &                    &+50.98 & 0.34 &108 & Y & Cr 155\\
\object{$\lambda$ Ori}  &  9  &0.25 &  416 &7.10 &                    &+27.73 & 0.13 &285 & N & Collinder 69, Briceno 1, MWSC 0531\\
\object{Berkeley 31}    &  9  &0.35 & 7177 &9.45 &                    &+61.39 & 1.25 & 77 & Y & Biurakan 7\\
\object{Tombaugh 2}     &  9  &0.83 & 9316 &9.21 &                    &+122.47& 0.40 & 11 & Y & Haffner 2\\
\object{NGC 2324}       &  8  &0.40 & 4214 &8.73 &                    &+47.09 & 1.34 & 14 & Y & Mel 59\\ 
\object{NGC 6405}       &  8  &0.49 &  459 &7.54 &                    &-7.42  & 0.41 &124 & N & M38, M43, [KC2019] Theia 122 \\
\object{NGC 6791}       &  8  &0.70 & 4231 &9.80 &                    &-47.75 & 0.17 & 57 & N & Be 46\\
\object{ASCC 50}        &  8  &0.99 &  917 &7.06 &                    &+21.44 & 0.26 &158 & N & Alessi 43 \\
\object{NGC 6281}       &  7  &0.30 &  539 &8.71 &                    &-5.05  & 0.39 & 77 & Y & [KC2019] Theia 325\\
\object{NGC 6633}       &  7  &0.30 &  424 &8.84 &2.709$^7$           &-29.10 & 1.07 & 72 & Y & [KC2019] Theia 924\\
\object{Berkeley 22}    &  7  &1.99 & 6225 &9.39 &                    &+94.31 & 0.44 & 17 & N &\\
\object{Berkeley 44}    &  7  &2.75 & 2863 &9.16 &                    &-7.54  & 0.58 & 41 & N &\\
\object{Czernik 30}     &  7  &0.62 & 6647 &9.46 &                    &+82.07 & 0.73 & 46 & Y &\\
\object{NGC 4337}       &  7  &1.06 & 2450 &9.16 &3.021$^2$           &-17.26 & 0.29 & 19 & Y &\\
\object{NGC 4815}       &  6  &1.75 & 3295 &8.57 &                    &-28.11 & 1.09 & 83 & Y &\\
\object{Pismis 15}      &  6  &1.89 & 2599 &8.94 &2.744$^2$           &+34.22 & 0.94 & 53 & Y &\\
\object{Pismis 18}      &  6  &1.81 & 2860 &8.76 &                    &-24.85 & 0.89 & 42 & Y & IC 4291\\
\object{Berkeley 36}    &  6  &1.42 & 4360 &9.83 &                    &+62.94 & 0.25 &112 & Y &\\
\object{Czernik 24}     &  6  &1.63 & 3981 &9.43 &                    &+21.52 & 0.48 & 71 & Y &\\
\object{Collinder 261}  &  6  &0.81 & 2850 &9.80 &                    &-24.43 & 0.17 &230 & Y & Harward 6\\
\object{NGC 2232}       &  5  &0.01 &  315 &7.25 &2.356$^2$,2.281$^1$ &+25.75 & 0.11 & 69 & Y & [KC2019] Theia 55\\
\object{NGC 2660}       &  5  &1.19 & 2788 &8.97 &                    &+22.47 & 0.27 &  8 & Y & Mel 92\\
\object{Berkeley 25}    &  5  &1.07 & 6780 &9.39 &                    &+108.07& 9.19 & 14 & N &\\
\object{Ruprecht 7}     &  5  &1.75 & 5851 &8.37 &                    &+77.38 & 0.32 &  7 & Y & Berkeley 33\\
\object{25 Ori}         &  5  &0.25 &  416 &7.10 &                    &+27.73 & 0.13 &285 & N & Collinder 69\\
\object{Collinder 110}  &  5  &1.14 & 2183 &9.26 &                    &+38.15 & 0.21 & 48 & Y &\\
\object{Ruprecht 147}   &  4  &0.06 &  323 &9.48 &                    &+42.18 & 0.38 & 99 & Y & NGC 6774, [KC2019] Theia 1531\\
\object{NGC 6192}       &  4  &1.57 & 1737 &8.38 &3.265$^2$           &-7.71  & 0.19 &  6 & Y & Cr 309\\
\object{NGC 6709}       &  4  &0.72 & 1041 &8.28 &2.832$^2$           &-4.22  & 1.09 & 91 & Y & [KC2019] Theia 985\\
\object{$\rho$ Oph}     &  4  &     &  139 &5.5-6.8 &                 &       &      &     & N & \\
\object{ESO092-05}      &  4  &0.17 &12444 &9.65 &                    &+57.40 & 3.38 & 22 & N & \\ 
\object{Berkeley 30}    &  4  &1.27 & 5383 &8.47 &2.959$^2$           &+46.94 & 0.84 & 53 & Y & Biurakan 9\\
\object{Berkeley 73}    &  4  &0.69 & 6158 &9.15 &                    &+97.51 & 0.53 & 28 & N &\\
\object{Berkeley 75}    &  4  &0.29 & 8304 &9.23 &                    &+122.41& 1.73 & 23 & Y & ESO490-50\\
\object{NGC 3293}       &  3  &0.90 & 2710 &7.01 &3.296$^2$           &-7.04  & 3.37 & 25 & N & Cr 224, Mel 100\\
\object{NGC 3766}       &  3  &0.66 & 2123 &7.36 &                    &-16.18 & 0.59 & 21 & N & Cr 248\\   
\object{NGC 6583}       &  3  &1.52 & 2053 &9.08 &                    &-1.43  & 0.49 & 34 & Y &\\ 
\object{$\gamma$ Vel}   &  3  &     &  330 &7.30 &                    &       &      &    & N &\\
\object{NGC 5822}       &  2  &0.39 &  854 &8.96 &                    &-28.71 & 0.14 & 39 & N & Mel 130,[KC2019] Theia 1174 \\
\object{Berkeley 20}    &  2  &0.37 & 8728 &9.68 &                    &+75.65 & 1.38 &  3 & Y &\\
\object{Berkeley 29}    &  2  &0.24 &12604 &9.49 &                    &+25.72 & 1.99 &  9 & Y &\\
\object{Ruprecht 4}     &  2  &1.24 & 4087 &8.93 &2.868$^2$           &       &       &    & Y &  \\
\object{Collinder 197}  &  2  &1.42 &  955 &7.15 &                    &+28.84 & 0.50 &101 & N & ESO313-13, [KC2019] Theia 28\\
\object{NGC 2244}       &  1  &1.46 & 1478 &7.10 &                    &+32.86 & 1.00 &142 & N & NGC 2239\\
\object{NGC 6404}       &  1  &3.47 & 2500 &8.00 &                    &+10.12 & 0.26 &  9 & N &\\
\object{NGC 6649}       &  1  &3.90 & 2124 &7.85 &                    &+4.50  & 1.09 & 86 & N &\\
\hline
\multicolumn{11}{l}{Notes: $^1$ \citet{ebrahimi22}; $^2$ \citet{almeida2023} }\\
\label{infoA1}
\end{longtable}

\onecolumn
\begin{landscape}
\centering

\begin{table*}
\setlength{\tabcolsep}{1.mm}

\caption{Properties of the individual stars (excerpt).}
\begin{tabular}{lllllllllllllllll}
\hline
Cluster     & DR3 ID              & object            & RV    & err  &T$_{eff}$ &err &$\log~g$ &err &[Fe/H] &err & Na1   &err      & O1   &err      &RV      &err \\
            & Gaia                & GES               & GES   &      &GES       &    &GES      &    &GES    &    & GES   &        & GES  &   &Gaia    &    \\
\hline
25 Ori & 3222177030594592256 & 05233297+0135176  &  14.41 &  0.37 & 6568 &  39 & 4.06 & 0.07 & -0.07 & 0.05 &  6.24 & 0.14 &      &      &  17.27 & 1.84 \\ 
     lamOri & 3339510731055206528 & 05352469+1011453  &  29.33 &  0.37 & 6200 &  37 & 4.16 & 0.07 & -0.06 & 0.05 &  6.12 & 0.05 &      &      &  29.67 & 0.72 \\ 
     lamOri & 3337936092965373568 & 05361858+0945089  &  29.06 &  0.37 & 5040 &  35 & 4.42 & 0.05 & -0.10 & 0.05 &  6.33 & 0.14 &      &      &  27.62 & 4.73 \\ 
     lamOri & 3336149489649827840 & 05432474+0906084  &  25.01 &  0.37 & 4542 &  30 & 4.10 & 0.05 & -0.18 & 0.05 &  5.96 & 0.05 &      &      &  23.95 & 15.75\\ 
    Blanco1 & 2320833372790628480 & 00013320-3012597  &   6.59 &  0.37 & 5311 &  31 & 4.55 & 0.05 & -0.05 & 0.05 &  6.03 & 0.02 &      &      &   7.33 & 1.11 \\ 
    Blanco1 & 2333005344467248768 & 00024879-2918422  &   5.69 &  0.37 & 5446 &  31 & 4.42 & 0.05 & -0.02 & 0.05 &  6.08 & 0.02 &      &      &   5.03 & 1.07 \\ 
    Blanco1 & 2320795340855502336 & 00023546-3007019  &  -3.27 &  0.37 & 5303 &  30 & 4.50 & 0.05 & -0.09 & 0.05 &  5.91 & 0.05 &      &      &   7.06 & 3.17 \\ 
    Blanco1 & 2320795237776344704 & 00030028-3003216  &   6.52 &  0.37 & 5017 &  31 & 4.48 & 0.05 & -0.07 & 0.05 &  6.19 & 0.02 &      &      &   5.94 & 1.55 \\ 
    Blanco1 & 2320869901488170624 & 00032061-2949227  &   5.58 &  0.37 & 6230 &  32 & 4.36 & 0.05 & -0.05 & 0.05 &  6.19 & 0.06 &      &      &   5.67 & 0.61 \\ 
    Blanco1 & 2320591205353715712 & 00050824-3029421  &   6.04 &  0.37 & 5754 &  32 & 4.42 & 0.05 &  0.01 & 0.05 &  6.15 & 0.02 &      &      &   5.36 & 0.58 \\ 
    Blanco1 & 2320925564263593088 & 00043317-2938281  &   5.47 &  0.37 & 6317 &  33 & 4.18 & 0.05 & -0.03 & 0.05 &  6.09 & 0.05 &      &      &   5.81 & 0.64 \\ 
    Blanco1 & 2320916768170630912 & 00055905-2939046  &   5.41 &  0.37 & 5923 &  36 & 4.47 & 0.05 & -0.01 & 0.05 &  6.11 & 0.14 &      &      &   5.66 & 1.43 \\ 
    Blanco1 & 2320617529209321728 & 00045884-3009416  &   5.93 &  0.37 & 6038 &  44 & 4.34 & 0.09 & -0.12 & 0.08 &  6.20 & 0.12 &      &      &   5.77 & 0.67 \\ 
    Blanco1 & 2320617116892496256 & 00052902-3008321  &   8.58 &  0.37 & 4832 &  30 & 4.58 & 0.05 & -0.07 & 0.05 &  6.07 & 0.02 & 9.00 & 0.10 &   5.84 & 1.96 \\ 
    Blanco1 & 2320615983021130752 & 00055472-3006258  &   6.02 &  0.37 & 5677 &  30 & 4.47 & 0.05 & -0.01 & 0.05 &  6.12 & 0.05 &      &      &       &      \\ 
       Be20 & 3221067898239847168 & 05323677+0011048  &  78.81 &  0.37 & 4847 &  31 & 2.70 & 0.05 & -0.32 & 0.06 &  6.04 & 0.02 & 8.66 & 0.15 &       &      \\ 
       Be20 & 3221067902536353152 & 05323896+0011203  &  78.91 &  0.37 & 4382 &  31 & 1.81 & 0.06 & -0.43 & 0.06 &  5.92 & 0.05 & 8.54 & 0.10 &  83.45 & 3.97 \\ 
       Be21 & 3424170411975583616 & 05513844+2147197  &   0.72 &  0.37 & 4368 &  33 & 1.80 & 0.06 & -0.26 & 0.04 &  5.91 & 0.02 & 8.61 & 0.12 &   1.25 & 2.82 \\ 
       Be21 & 3424170549414520832 & 05514200+2148497  &  -0.46 &  0.37 & 4520 &  33 & 2.25 & 0.05 & -0.18 & 0.04 &  6.09 & 0.02 & 8.78 & 0.15 &  -4.21 & 4.44 \\ 
       Be21 & 3424170515054788736 & 05514204+2148027  &   0.91 &  0.37 & 4509 &  31 & 2.15 & 0.05 & -0.23 & 0.05 &  5.98 & 0.05 & 8.71 & 0.15 &   1.66 & 3.21 \\ 
 ....       & & & & & & & & & & & & & & & &\\ 
 Trumpler23 & 5980830577767550976 & 16010639-5331056  & -62.41 &  0.37 & 4725 &  33 & 2.63 & 0.05 &  0.15 & 0.04 &  6.52 & 0.05 & 8.82 & 0.07 & -63.63 & 1.06 \\ 
 Trumpler23 & 5980841916481958144 & 16003935-5332367  & -61.36 &  0.37 & 4725 &  33 & 2.63 & 0.05 &  0.18 & 0.04 &  6.57 & 0.05 & 8.93 & 0.05 & -60.93 & 0.70 \\ 
 Trumpler23 & 5980842431878045696 & 16004312-5330509  & -62.22 &  0.37 & 4815 &  33 & 2.85 & 0.05 &  0.24 & 0.04 &  6.53 & 0.05 & 8.71 & 0.07 & -63.99 & 2.32 \\ 
 Trumpler23 & 5980830607806894720 & 16005168-5332013  & -63.26 &  0.37 & 4784 &  33 & 2.60 & 0.05 &  0.21 & 0.04 &  6.57 & 0.05 & 8.88 & 0.05 & -62.63 & 0.83 \\ 
 Trumpler23 & 5980830405968813824 & 16010025-5333101  & -60.66 &  0.37 & 4805 &  33 & 2.72 & 0.05 &  0.20 & 0.04 &  6.55 & 0.05 & 8.74 & 0.07 & -62.85 & 1.31 \\ 
 Trumpler23 & 5980832055236335744 & 16010770-5329374  & -61.94 &  0.37 & 4753 &  32 & 2.63 & 0.05 &  0.22 & 0.04 &  6.57 & 0.05 & 8.85 & 0.06 & -63.70 & 0.83 \\ 
 Trumpler23 & 5980842500597538304 & 16004025-5329439  & -56.95 &  0.37 & 4765 &  33 & 2.68 & 0.05 &  0.19 & 0.04 &  6.56 & 0.05 &      &      & -58.23 & 2.67 \\ 
 Trumpler23 & 5980841916481954432 & 16004035-5333047  & -69.15 &  0.37 & 4768 &  33 & 2.96 & 0.05 &  0.03 & 0.04 &  6.27 & 0.04 &      &      & -65.34 & 0.82 \\ 
 Trumpler23 & 5980830199810393216 & 16005220-5333362  & -62.92 &  0.37 & 4817 &  32 & 2.65 & 0.05 &  0.26 & 0.05 &  6.62 & 0.05 & 8.93 & 0.04 & -61.41 & 3.03 \\ 
 Trumpler23 & 5980830646487029248 & 16005798-5331476  & -60.43 &  0.37 & 4767 &  32 & 2.70 & 0.05 &  0.23 & 0.04 &  6.54 & 0.05 & 8.92 & 0.04 & -58.49 & 1.17 \\ 
 \hline
\end{tabular}

\label{infoA2}
\end{table*}

\end{landscape}

\begin{table*}
\caption{Candidate binaries on the basis of the RV difference between $Gaia$
DR3 and GES.}
\setlength{\tabcolsep}{1.mm}
\begin{tabular}{lrrrrrrrrrccr}
\hline\hline
Cluster    & Gaia DR3 ID        &RV    &err    &num &Vbroad &err    &RUWE    &RV    &err &proba &mem3d &$\Delta$RV \\
           &                    &(GDR3)&       &    &(GDR3) &       &        &    &(GES) &       &      &          \\  
\hline
  Blanco 1 &2320795340855502336 &  7.06 & 3.17  & 19&       &       &  7.699 & -3.270 &0.37 &         &0.95 &-10.33\\
     Be 21 &3424170549414515200 &  3.62 & 1.89  & 14&       &       &  1.074 & -3.430 &0.38 &0.8 &1.00 & -7.05\\
     Be 31 &3157239843797500672 & 75.13 & 8.60  &  8&       &       &  0.995 & 56.600 &0.37 &0.9 &1.00 &-18.53\\
     Be 36 &3032952217629397760 & 55.90 & 6.66  & 10&       &       &  1.056 & 63.920 &0.37 &1.0 &1.00 &  8.02\\
     Be 73 &3007967052833159296 & 81.90 & 7.39  & 20&       &       &  1.019 & 96.790 &0.37 &1.0 &1.00 & 14.89\\
     Be 81 &4265584329578299904 & 39.32 & 4.44  & 12&       &       &  0.952 & 48.900 &0.37 &1.0 &1.00 &  9.58\\
     Be 81 &4265584054700298240 & 37.80 & 9.06  &  9&       &       &  1.010 & 48.040 &0.37 &0.9 &1.00 & 10.24\\
     Be 81 &4265582577231602304 & 30.92 &12.62  &  6&       &       &  0.956 & 48.440 &0.37 &1.0 &1.00 & 17.52\\
     Be 81 &4265582405432857472 & 60.34 & 5.52  &  5&       &       &  1.090 & 48.140 &0.37 &1.0 &1.00 &-12.20\\
     Cha I &5225317513655541504 & -9.76 & 6.78  & 25&       &       & 25.026 & 14.430 &0.37 &         &1.00 & 24.19\\
     Cha I &5201154444261256704 & 71.36 &10.71  &  6&       &       &  1.075 & 15.150 &0.37 &         &1.00 &-56.21\\
     Cha I &5201362423758639744 & -6.59 &13.19  & 16&       &       &  1.136 & 15.140 &0.38 &         &1.00 & 21.73\\
   IC 4665 &4474066401451091840 &  6.59 & 7.70  & 19& 39.42 & 29.36 &  0.939 & 17.590 &0.37 &1.0 &0.95 & 11.00\\
      M 67 & 604911410242410752 & 38.82 & 5.29  & 24&       &       & 15.870 & 50.880 &0.37 &         &0.96 & 12.06\\
      M 67 & 604922985178465152 & 34.60 & 2.26  & 25&       &       &  1.373 & 49.260 &0.37 &1.0 &0.99 & 14.66\\
      M 67 & 604917629355038848 & 37.05 & 9.14  & 19&       &       &  0.918 &-31.680 &0.37 &0.8 &1.00 &-68.73\\
      M 67 & 604917491916095872 & 52.81 & 9.52  & 19& 16.31 & 20.02 &  0.942 & 14.400 &0.37 &0.8 &0.99 &-38.41\\
    Mel 71 &3033961638024647808 & 65.42 & 4.96  & 22&       &       &  1.020 & 84.120 &0.37 &0.9 &     & 18.70\\
    Mel 71 &3033962050339630208 & 61.01 & 5.51  & 21&  7.74 &  8.05 &  1.028 & 32.640 &0.37 &1.0 &     &-28.37\\
    Mel 71 &3033958747506151424 & 53.10 & 3.87  & 26& 10.56 & 17.82 &  0.941 & 45.700 &0.37 &1.0 &     & -7.40\\
    Mel 71 &3033962187778553600 & 54.50 & 2.98  & 20&  6.95 & 12.44 &  1.482 & 70.300 &0.37 &0.7 &     & 15.80\\
  NGC 2264 &3326929191297621120 & -7.66 & 8.03  &  8&       &       &  1.152 & 32.940 &0.37 &0.9 &0.98 & 40.60\\
  NGC 2264 &3326904521005483136 & 33.12 & 4.64  &  9&       &       &  0.921 & 25.160 &0.37 &0.9 &0.98 & -7.96\\
  NGC 2264 &3326685443313414144 &-22.21 & 5.52  &  8&       &       &  1.088 & 18.800 &0.38 &         &0.97 & 41.01\\
  NGC 2264 &3326696124896220928 &-27.97 & 4.79  &  9&       &       &  1.021 & 21.050 &0.27 &         &1.00 & 49.02\\
  NGC 2420 & 865398496685953536 & 84.54 & 9.19  &  9&       &       &  1.001 & 74.350 &0.10 &0.9 &1.00 &-10.19\\
  NGC 2451 &5538749417871537024 & 13.15 & 2.23  & 21& 37.49 & 27.60 &  1.338 & -2.720 &0.37 &0.9 &     &-15.87\\
  NGC 2451 &5538817690669417984 &  7.20 & 4.76  & 16&       &       &  1.649 & 14.610 &0.38 &0.8 &     &  7.41\\
  NGC 2516 &5290767115830162560 & 19.13 & 7.48  & 20&       &       &  2.364 & 49.630 &0.38 &0.7 &0.97 & 30.50\\
  NGC 2547 &5514369229297396736 &-24.70 &10.25  & 10&       &       &  0.994 & 13.520 &0.37 &0.9 &1.00 & 38.22\\
  NGC 2547 &5514371874997910656 & 33.46 & 5.67  &  7&       &       &  4.536 & 13.640 &0.37 &         &1.00 &-19.82\\
  NGC 3532 &5340146079993296384 &  6.88 & 3.03  & 15&       &       &  2.920 & -7.780 &0.37 &         &0.97 &-14.66\\
  NGC 3532 &5340215349223412992 & 27.12 & 5.36  &  7&       &       &  1.053 & 37.150 &0.37 &1.0 &0.85 & 10.03\\
  NGC 6253 &5935992940300723072 &-23.17 &13.83  & 17&       &       &  1.006 &-30.630 &0.37 &0.8 &1.00 & -7.46\\
  NGC 6253 &5935943530992746880 &-45.56 &12.08  & 12&       &       &  0.976 &-29.260 &0.37 &0.8 &1.00 & 16.30\\
  NGC 6253 &5935945180260232064 &-12.45 &17.47  & 10&       &       &  1.010 &-27.670 &0.37 &         &0.99 &-15.22\\
  NGC 6405 &4054223731895900288 &-53.44 & 4.78  & 15&       &       &  0.855 & -8.610 &0.37 &1.0 &0.99 & 44.83\\
  NGC 6530 &4066064956700837248 & -3.37 & 7.63  & 15&323.50 & 49.20 &  0.789 &-15.510 &1.60 &         &0.92 &-12.14\\
 Pismis 15 &5410176380718863232 & 23.13 & 2.53  & 24&       &       &  1.034 & 30.680 &0.37 &1.0 &0.99 &  7.55\\
   Rup 134 &4056457630330166784 &-38.39 &10.15  &  2&       &       &  1.434 &-45.950 &0.37 &         &0.96 & -7.56\\
Trumpler 5 &3326785773750746112 & 59.96 & 4.47  &  7&       &       &  1.040 & 51.130 &0.37 &1.0 &1.00 & -8.83\\
Trumpler 5 &3326786564024669696 & 58.16 & 5.52  & 10&       &       &  0.957 & 51.340 &0.37 &1.0 &1.00 & -6.82\\
 gamma Vel &5519267038205288832 & 14.89 & 4.90  & 15& 76.91 & 35.31 &  0.998 &  1.720 &0.37 &         &0.88 &-13.17\\
\hline
\end{tabular}
\label{t:deltaRV}
\end{table*}

\end{appendix}

\end{document}